\documentstyle[12pt]{article}
\begin{document}
\begin{center}
{\bf{{The Ehrenfest Theorem in Quantum Field Theory}}} 

\vspace{1.0cm}

R.Parthasarathy {\footnote{e-mail address: sarathy@cmi.ac.in}} \\
Chennai Mathematical Institute \\
H1, SIPCOT IT Park, Padur Post \\
Siruseri 603103, India. \\
\end{center}

\vspace{1.0cm}

{\noindent{\it{Abstract}}}

\vspace{0.5cm}

The validity of the Ehrenfest's theorem in Abelian and non-Abelian quantum field theories is examined. The 
gauge symmetries are taken to be unbroken. By suitably choosing the physical subspace, the above 
validity is proven in both the cases.

\vspace{1.0cm}

{\noindent{\bf{Key words:}} Ehrenfest's theorem - Schr\"{o}dinger equation - expectation values - 
Dirac equation - Abelian field theory - gauge fixing - quantum lagrangian - physical subspace - 
Non-Abelian field theory - Quantum Chromo Dynamics - path integral approach - Faddeev-Popov ghosts - 
quantum equations - BRS transformation - Global gauge and scale transformations - physical states.

\vspace{0.5cm}


\vspace{1.5cm} 

\begin{center}
{\bf(Dedicated to the memory of Prof.Alladi Ramakrishnan)}
\end{center}

\vspace{1.0cm}

Professor Alladi Ramakrishnan founded the Institute of Mathematical Sciences (MATSCIENCE) in 1962 and 
attracted bright young students interested in theoretical physics. His contributions to the theory of 
Stochastic processes, elementary particle physics and Generalized Clifford Algebras will be remembered 
forever. He was instrumental in my joining MATSCIENCE in 1977 and encouraged me till his end in my 
research work. I consider it my duty to dedicate this article in his memory. 

\vspace{1.0cm}

{\noindent{\bf{1. Quantum Mechanics}}}

\vspace{0.5cm}

In quantum mechanics, it is reasonable to expect the motion of a wave packet to agree with the motion of 
the corresponding classical particle whenever the potential energy changes by a small amount over the 
dimensions of the wave packet. If we mean by the 'position' and the 'momentum' vectors of the wave 
packet, their expectation values, then we can show that the classical and the quantum motions agree. This 
important result is known as Ehrenfest's theorem [1,2]. To illustrate this theorem, let us first consider 
non-relativistic quantum mechanics. We have the Schr\"{o}dinger equation 
\begin{eqnarray}
i\hbar \frac{\partial \psi(\vec{x},t)}{\partial t}
&=&-\frac{ {\hbar}^2}{2m}\ {\vec{\nabla}}^{\ 2}\psi(\vec{x},t) + V(\vec{x}\ )\psi(\vec{x},t), 
\nonumber \\
-i\hbar \frac{\partial {\psi(\vec{x},t)}^{\dagger}}{\partial t}
&=&-\frac{ {\hbar}^2}{2m}\ {\vec{\nabla}}^{\ 2}
{\psi(\vec{x},t)}^{\dagger}+V(\vec{x}\ ){\psi(\vec{x},t)}^{\dagger},
\end{eqnarray}
where $m$ is the mass of the particle and $V(\vec{x}\ )$ is the real potential. 

\vspace{0.5cm}

We shall take the wave function $\psi(\vec{x},t)$ in (1) as normalized. Then the expectation value of the 
$x$-component of the position operator and its time derivative are 
\begin{eqnarray}
\langle x\rangle &=&\int {\psi}^{\dagger}\ x\ \psi\ d\tau, \nonumber \\
\frac{d}{dt}\langle x\rangle &=&\int \Big(\frac{d{\psi}^{\dagger}}{dt}\Big) x \psi d\tau + \int 
{\psi}^{\dagger}x \Big(\frac{d\psi}{dt}\Big) d\tau.
\end{eqnarray}
Using (1), it follows 
\begin{eqnarray}
\frac{d}{dt}\langle x\rangle &=&-\frac{i\hbar}{m}\int {\psi}^{\dagger}\frac{\partial }{\partial x}
\psi d\tau \ =\ \frac{1}{m}\langle p_x\rangle.
\end{eqnarray}
Similarly starting from $\langle p_x\rangle = -i\hbar \int {\psi}^{\dagger}\frac{\partial}{\partial x}
\psi d\tau$, it is easy to find 
\begin{eqnarray}
\frac{d}{dt}\langle p_x\rangle &=& \langle -\frac{\partial V(\vec{x}\ )}{\partial x}\rangle.
\end{eqnarray}
 
\vspace{0.5cm}

From (2) and (4), we note that the {\it{classical}} equations of motion 
\begin{eqnarray}
\frac{d\vec{x}}{dt}\ =\ \frac{\vec{p}}{m}&;& \frac{d\vec{p}}{dt}\ =\ -\vec{\nabla}V(\vec{x}\ ),
\end{eqnarray}
are satisfied by their expectation values in quantum mechanics. The wave packet moves like a classical 
particle whenever the expectation value gives a good representation of the classical variable. They 
provide an example of the correspondence principle [1,2]. 

\vspace{0.5cm}

In the case of relativistic quantum mechanics, the manipulations are a little less direct. 
We consider the Dirac equation [3].  
\begin{eqnarray}
H&=&\vec{\alpha}\cdot \vec{p}+\beta m , \nonumber \\
i\hbar \frac{\partial \psi}{\partial t}&=&(\vec{\alpha}\cdot\vec{p}+\beta m )\psi,
\end{eqnarray} 
where $\vec{\alpha}$ and $\beta$ are hermitian $4\times 4$ matrices and $\psi$ is a $4\times 1$ column 
vector. We shall set the velocity of light $c$ to unity hereafter.  
By using the Heisenberg equation of motion $\frac{dx}{dt}=\frac{1}{i\hbar}[x,H]$, it is seen that 
\begin{eqnarray}
\int {\psi}^{\dagger}\ \frac{dx}{dt}\ \psi d\tau &=& \int {\psi}^{\dagger} {\alpha}_x \psi d\tau.
\end{eqnarray}
First, we recall the plane wave solutions ${\psi}^{(i)}$ [3] of the Dirac equation, 
\begin{eqnarray}
{\psi}^1(x)=\sqrt{\frac{E+m}{2m}}\ e^{-ipx}\left(\begin{array}{c}
1 \\
0 \\
\frac{p_z}{(E+m)} \\
\frac{p_+}{(E+m)} \\
\end{array}\right)&;&{\psi}^2(x)=\sqrt{\frac{E+m}{2m}}\ e^{-ipx}\left(\begin{array}{c}
0 \\
1 \\
\frac{p_-}{(E+m)} \\
-\frac{p_z}{(E+m)} \\
\end{array}\right), \nonumber \\
{\psi}^3(x)=\sqrt{\frac{E+m}{2m}}\ e^{ipx}\left(\begin{array}{c}
\frac{p_z}{(E+m)} \\
\frac{p_+}{(E+m)} \\
1 \\
0 \\
\end{array}\right) &;& {\psi}^4(x)=\sqrt{\frac{E+m}{2m}}\ e^{ipx}\left(\begin{array}{c}
\frac{p_-}{(E+m)} \\
-\frac{p_z}{(E+m)} \\
0 \\
1 \\
\end{array}\right), \nonumber 
\end{eqnarray} 
corresponding to positive energy $(E>0)$ spin-up, spin-down states and negative energy $(E<0)$ 
spin-up, spin-down states of the electron respectively, $px=Et-\vec{p}\cdot \vec{x}$ and 
$p_{\pm}=p_x\pm ip_y$. These solutions 
satisfy ${\psi^{(i)}}^{\dagger}(x){\psi}^{(j)}(x)\ =\ \frac{E}{m}{\delta}^{i,j}$ ($i,j=1,2,3,4$). 
Using them, we construct the wave packets
\begin{eqnarray}
\Psi(E>0)&=&\sum_{i=1}^2\int A_i(\vec{p}\ ){\psi}^{(i)} d^3p, \nonumber \\
\Psi(E<0)&=&\sum_{i=3}^4\int A_i(\vec{p}\ ){\psi}^{(i)} d^3p,
\end{eqnarray}
and note 
\begin{eqnarray}
\int {\Psi}^{\dagger}(E>0)\Psi(E>0) d^3x&=&\int d^3p\frac{E}{m}\{|A_1(\vec{p}\ )|^2+
|A_2(\vec{p}\ )|^2\}.
\end{eqnarray}
Similar expression can be written for $\Psi(E<0)$. Using the explicit representation of the 
${\alpha}_x$ matrix [3], we have 
\begin{eqnarray}
\int {\Psi}^{\dagger}(E>0){\alpha}_x\Psi(E>0) d^3x&=&\int d^3p \Big(\frac{p_x}{m}\Big) \{
|A_1(\vec{p}\ )|^2+|A_2(\vec{p}\ )|^2\}.
\end{eqnarray}
From (7), (9) and (10), it follows $\frac{d}{dt}\langle x\rangle = \langle \frac{p_x}{m}\rangle$, 
showing the validity of the Ehrenfest's theorem. Further, we consider the Dirac particle in an external 
electromagnetic field. Setting the vector potential zero (for simplicity), the Dirac hamiltonian is 
\begin{eqnarray}
H&=&\vec{\alpha}\cdot \vec{p}+\beta m -e\phi, 
\end{eqnarray}
where $\phi$ is the scalar potential. Using the Heisenberg equation of motion for a dynamical 
variable $F$, $\frac{dF}{dt}=\frac{1}{i\hbar}[F,H]$, 
it follows that $\frac{d\vec{p}}{dt}=-\vec{\nabla}(-e\phi)$ and so $\langle \frac{d\vec{p}}{dt}
\rangle =-\langle \vec{\nabla}(-e\phi)\rangle$, showing the validity of the Ehrenfest's theorem. 

\vspace{0.5cm}

Thus in quantum mechanics, we see that the expectation values of the position and the momentum 
operators satisfy the classical equations of motion. We would like to extend this to quantum field theory. 

\vspace{1.0cm}

{\noindent{\bf{2. Abelian field theory}}}

\vspace{0.5cm}

We consider the lagrangian density for the electromagnetic field minimally coupled to a source 
$j^{\mu}(x)$ (Dirac current)
\begin{eqnarray}
{\cal{L}}&=&-\frac{1}{4}F_{\mu\nu}F^{\mu\nu} + eA_{\mu}(x)j^{\mu}(x),
\end{eqnarray}
where $A_{\mu}(x)$ is the electromagnetic field, $e$ is the coupling strength and 
\begin{eqnarray} 
F_{\mu\nu}&=&{\partial}_{\mu}A_{\nu}-{\partial}_{\nu}
A_{\mu}. 
\end{eqnarray}
The corresponding {\it{classical equations}} (Euler-Lagrange equations) are 
\begin{eqnarray}
{\partial}_{\mu}F^{\mu\nu}+ej^{\nu}&=&0. 
\end{eqnarray}
Eqn.14 is the {\it{classical equation of motion}} and gives the Maxwell equations with source. 

\vspace{0.5cm}

It is well known that the manifestly covariant theory of 
massless vector field is to be quantized with indefinite metric [4]. The impossibility of quantizing 
the electromagnetic field with positive definite metric has been shown by Mathews, Seetharaman and Simon 
[5]. A physically meaningful theory is constructed by introducing a 'subsidiary condition', which is a 
condition defining the {\it{physical subspace}} of the indefinite metric Hilbert space of the 
electromagnetic field. In here, we 
follow the $B$-field formalism of Nakanishi [6]. In order to quantize the above lagrangian, one has to 
fix the gauge. This is seen by considering  
the coefficient of the terms quadratic in $A_{\mu}$ in the action $S=\int d^4x 
{\cal{L}}$ (after a partial integration). This coefficient is the differential operator 
$\Box g^{\mu\nu}-{\partial}^{\mu}{\partial}^{\nu}$. The two-point function $\langle A_{\mu}(x)A_{\nu}(y)
\rangle$ is governed by the above differential operator.

\vspace{0.5cm}

The Feynman propagator for the photon (quantized electromagnetic field) is the inverse of this 
differential operator in the momentum space. As this differential operator is not invertible, the 
photon propagator is not defined. This difficulty is avoided by choosing a gauge. We choose the 
covariant gauge ${\partial}^{\mu}A_{\mu}=0$ and implement this gauge fixing in the lagrangian by adding 
the 'gauge fixing term' $-\frac{1}{2a}({\partial}^{\mu}A_{\mu})^2$ where $a$ is a parameter. This 
modifies the coefficient of the terms quadratic in $A_{\mu}$ in the action $S$ as $\Box g^{\mu\nu}
-{\partial}^{\mu}{\partial}^{\nu}+\frac{1}{a}{\partial}^{\mu}{\partial}^{\nu}$. This, in the 
momentum space is $-p^2g^{\mu\nu}+(1-\frac{1}{a})p^{\mu}p^{\nu}$ whose inverse is $-\frac{1}{p^2}
\{g_{\mu\nu}+\frac{a-1}{p^2}p_{\mu}p_{\nu}\}$, which is the Feynman propagator for the photon in the 
covariant gauge. 

\vspace{0.5cm}

We introduce the above covariant 
gauge fixing via $B(x)$, an auxiliary hermitian scalar field and consider the {\it{quantum lagrangian}} 
\begin{eqnarray}
{\cal{L}}&=&-\frac{1}{4}F_{\mu\nu}F^{\mu\nu}+B(x){\partial}^{\mu}A_{\mu}+\frac{a}{2}B^2(x) + 
eA_{\mu}(x)j^{\mu}(x),
\end{eqnarray}
where $a$ is a parameter. It is important to realize that the gauge field $A_{\mu}(x)$ and $B(x)$ in (15) 
are {\it{operators}} while the gauge field in (12) is a classical field. The quantum equations of motion 
from (15) are 
\begin{eqnarray}
{\partial}_{\mu}F^{\mu\nu}-{\partial}^{\nu}B(x)&=&-ej^{\nu}, \nonumber \\
{\partial}^{\mu}A_{\mu}+aB(x)&=&0.
\end{eqnarray}
Using the second equation to eliminate the $B$-field in the lagrangian, we recover the gauge fixing term 
$-\frac{1}{2a}({\partial}^{\mu}A_{\mu})^2$. By taking ${\partial}_{\nu}$ of the first equation and 
using the conservation of the current $j^{\nu}(x)$, namely ${\partial}_{\nu}j^{\nu}(x)=0$, 
we see that $B(x)$ satisfies the equation of motion for a massless scalar field, admitting positive 
and negative frequency solutions. Eqn.16 can be 
considered to be the {\it{quantum Maxwell equations}} while (14) is the classical equation of motion. The 
fields in (16) are operators and act on functions (states) in the indefinite metric Hilbert space. 
For this reason, this method of quantization is called "operator method of quantization". In 
order to ensure that physically meaningful degrees of freedom only contribute (the longitudinal and the 
time-like photons are unphysical) to the observables, we impose Gupta's subsidiary condition on the 
photon states by 
\begin{eqnarray}
B^+(x)|\phi\rangle &=&0,
\end{eqnarray}
where the superscript $+$ denotes the positive frequency part of $B(x)$. The physical subspace in the 
indefinite metric Hilbert space is defined in (17). {\it{The physical subspace $V_{phys}$ is the 
totality of the states $|\phi\rangle$ satisfying (17).}} Now consider the expectation value of the 
quantum equations of motion (16) between {\it{physical states $|\phi\rangle $}} defined in (17). They are  
\begin{eqnarray}
\langle \phi|{\partial}_{\mu}F^{\mu\nu}-{\partial}^{\nu}B(x)+ej^{\nu}|\phi \rangle &=&0;\ \ 
|\phi\rangle \ \in \ V_{phys}, \nonumber \\
\langle \phi|{\partial}_{\mu}A^{\mu}+aB(x)|\phi \rangle &=&0. 
\end{eqnarray}
Using $B^-=(B^+)^{\dagger}$ and (17), (18) becomes 
\begin{eqnarray}
\langle \phi|{\partial}_{\mu}F^{\mu\nu}+ej^{\nu}|\phi \rangle &=&0; \ \forall \ |\phi \rangle \ \in \ 
V_{phys}, \nonumber \\
\langle \phi|{\partial}_{\mu}A^{\mu}|\phi \rangle &=&0. 
\end{eqnarray}
Comparing (19) with (14), we see that the expectation value of the quantum equation of motion taken 
with the states in the physical subspace reproduces the classical equations of motion, generalizing the 
Ehrenfest's theorem to Abelian quantum field theory. Since the classical equation of motion is linear in 
$A_{\mu}(x)$, one can separate the positive and negative frequency parts and then the second equation 
above gives ${\partial}^{\mu}A_{\mu}^+(x)|\phi \rangle =0$, subsidiary operator condition of Gupta. This 
feature is not shared by the non-Abelian theory as there the classical equation of motion for the 
non-Abelian gauge field is non-linear and a separation into positive and negative frequency parts is not 
possible. 

\vspace{1.0cm}

{\noindent{\bf{3. Non-Abelian Field Theory}}}

\vspace{0.5cm}

As an example, we consider $SU(3)$ gauge theory relevant to Quantum Chromo Dynamics (QCD), the gauge 
theory of the strong interactions of quarks. The {\it{classical}} lagrangian density is given by 
\begin{eqnarray}
{\cal{L}}_{YM}&=&-\frac{1}{4}F_{\mu\nu}^a F^{\mu\nu a} + gA_{\mu}^a\ j^{\mu a},
\end{eqnarray}
where $j^{\mu a}$ is the external source (color current of the quark), $\mu,\nu$'s are the Lorentz 
indices, $a,b,c$'s are the $SU(3)$ group indices, $g$ is the coupling strength, and 
\begin{eqnarray}
F_{\mu\nu}^a&=&{\partial}_{\mu}A_{\nu}^a-{\partial}_{\nu}A_{\mu}^a+gf^{abc}A_{\mu}^bA_{\nu}^c.
\end{eqnarray}
In above, $f^{abc}$'s are the structure constants of $SU(3)$ and $g$ is also the coupling strength of the 
self interaction of the non-Abelian gauge fields. The above lagrangian is gauge invariant. This can be 
verified by using the infinitesimal gauge transformation on the gauge field $A_{\mu}^a$, namely 
\begin{eqnarray}
A_{\mu}^a &\rightarrow & A_{\mu}^a+D_{\mu}^{ab}{\omega}^b, \ \ {\omega}^a\in SU(3), \nonumber \\
D_{\mu}^{ab}&=&{\partial}_{\mu}{\delta}^{ab}+gf^{acb}A_{\mu}^c. 
\end{eqnarray}
Consider the first term in the lagrangian. Then it is found, using the Jacobi identity 
\begin{eqnarray}
f^{bcd}f^{dae}+f^{cad}f^{dbe}+f^{abd}f^{dce}&=&0, 
\end{eqnarray}
that 
\begin{eqnarray}
{\delta}_{gauge}(F^{\mu\nu a}F_{\mu\nu}^a)&=&2gf^{acb}F^{\mu\nu a}F_{\mu\nu}^c\ {\omega}^b\ \equiv \ 0.
\end{eqnarray} 
The {\it{classical equations of motion}} from (20) are 
\begin{eqnarray}
D_{\mu}^{ab}F^{\mu\nu b}+gj^{\nu a}&=&0.
\end{eqnarray}
 
\vspace{0.5cm}

The operator $D_{\mu}^{ab}$ in (22) is called the covariant derivative in the adjoint representation and 
using the Jacobi identity (23), it is found that the commutator $[D_{\mu}, D_{\nu}]^{ab}=-gf^{abq}F_{
\mu\nu}^q$. Acting on (25) by $D_{\nu}^{ca}$, using the commutator, it is seen that
\begin{eqnarray}
D_{\nu}^{ab}j^{\nu b}&=&0,
\end{eqnarray}
that is, the current $j^{\nu a}$ is covariantly conserved. As the source $j^{\mu a}$ is gauge invariant,
in the action integral the second term in the lagrangian is invariant using (22) and (26) after one 
partial integration. Thus the lagrangian in (20) is gauge invariant. 

\vspace{0.5cm}

Using the covariant derivative, the classical equation of motion (22) can be rewritten as 
\begin{eqnarray}
{\partial}_{\mu}F^{\mu\nu a}+gf^{acb}A_{\mu}^c F^{\mu\nu b}+gj^{\nu a}&=&0, \nonumber \\
{\partial}_{\mu}F^{\mu\nu a}&=&-gJ^{\nu a},\ \ \ where \nonumber \\ 
J^{\nu a}&\equiv & j^{\nu a}+f^{acb}A_{\mu}^cF^{\mu\nu b}.
\end{eqnarray}
The current $J^{\nu a}$ contains besides the matter contribution, the non-Abelian fields. The 
non-Abelian fields themselves act as the source (like in gravity). By inspection, we see that 
${\partial}_{\nu}J^{\nu a}=0$, i.e., the current $J^{\nu a}$ is ordinarily conserved. 

\vspace{0.5cm}

An attempt to quantize (20) along the lines of the Abelian theory i.e., "operator method of 
quantization", runs into difficulty. The auxiliary 
fields $B^a(x)$ in this case do not satisfy $\Box B^a(x)=0$ due to the self-coupling property of the 
non-Abelian fields. So it is not possible to write down the positive and negative frequency parts. 
Further the classical equations of motion are non-linear. The proper method is to use the 
"path integral approach". For the reasons given in the Abelian field theory, here also we need to fix 
the gauge to obtain the propagator for the gauge fields $A_{\mu}^a(x)$. Further, in the "path 
integral method", one integrates all possible gauge field configurations. As the lagrangian (20) is 
gauge invariant, two gauge field configurations related by gauge transformation will give the same 
lagrangian. This, in the path integral amounts to double counting in the space of gauge fields. This 
is avoided by fixing the gauge and integrating over the space of gauge fields modulo gauge fixing. 
We choose the covariant gauge ${\cal{F}}^a={\partial}^{\mu}A_{\mu}^a(x)=0$. 

\vspace{0.5cm}

The above gauge fixing relation however does change by the gauge transformation and so the 
gauge variation of the gauge fixing relation is non-trivial in non-Abelian gauge theory. 
This, in the path integral approach, brings in the 
Faddeev-Popov ghost (anti-commuting scalars) fields. Using the results from the "path integral 
approach" [7], the lagrangian density for {\it{quantum}} non-Abelian theory can be written as 
\begin{eqnarray}
{\cal{L}}&=&-\frac{1}{4}F_{\mu\nu}^aF^{\mu\nu a}-{\partial}^{\mu}B^a\ A_{\mu}^a+\frac{\alpha}{2}B^aB^a 
-i{\partial}^{\mu}{\bar{c}}^{\ a}\ (D_{\mu}^{ab}c^b)+gj_{\nu}^aA^{\nu a},
\end{eqnarray}
where $\alpha$ is a gauge parameter and $c$'s are the ghost fields. They are hermitian 
\begin{eqnarray}
c^a=(c^a)^{\dagger}&;& {\bar{c}}^{\ a}=({\bar{c}}^{\ a})^{\dagger},
\end{eqnarray} 
and the ghost fields $c^a$ and ${\bar{c}}^{\ a}$ anti-commute. 

\vspace{0.5cm}

A comparison of (28) with (15) reveals that now we have (for $SU(3)$) eight auxiliary fields $B^a$ and 
a new term involving the Faddeev-Popov ghost fields. One can also quantize the Abelian massless field by 
the above procedure ("path integral approach") and in that case, the ghosts decouple from the gauge 
fields. In contrast, in (28), the 
fourth term contains coupling of the ghost fields with the gauge fields. This is crucial. 
The second and the third terms in (28) are the gauge fixing part and the fourth term is the 
Faddeev-Popov ghost part ${\cal{L}}_{FP}$. Using (29) and the anti-commuting property of the ghost fields, 
it is seen that ${\cal{L}}_{FP}^{\dagger}={\cal{L}}_{FP}$. The {\it{quantum equations of motion}} 
following from (28) are:
\begin{eqnarray}
D_{\mu}^{ab}F^{\mu\nu b}&=&{\partial}^{\nu}B^a-gj^{\nu a}-igf^{abc}({\partial}^{\nu}{\bar{c}}^{\ b})c^c, 
\nonumber \\
{\partial}_{\mu}A^{\mu a}+\alpha B^a&=&0, \nonumber \\
D_{\mu}^{ab}({\partial}^{\mu}{\bar{c}}^{\ b})&=&0, \nonumber \\
{\partial}_{\mu}(D^{\mu ab}c^b)&=&0.
\end{eqnarray}

\vspace{0.5cm}

Before considering the physical states, we recall that the quantum lagrangian (28) is gauge fixed. 
So, we do not have the local gauge invariance in (28). 
However, it was found by Becchi, Rouet and Stora (BRS) [8] that (28) is invariant under 
a special global transformation 
(First Global Transformation) involving Faddeev-Popov ghosts. This BRS transformation is given by 
\begin{eqnarray}
\delta A_{\mu}^a&=&D_{\mu}^{ab}c^b\ =\ [iQ,A_{\mu}^a], \nonumber \\ 
\delta \psi &=& igc^at^a\ \psi, \nonumber \\
\delta B^a&=&0\ =\ [iQ,B^a], \nonumber \\
\delta c^a&=&-\frac{g}{2}f^{abc}c^bc^c\ =\ \{iQ,c^a\}, \nonumber \\
\delta {\bar{c}}^a&=&iB^a\ =\ \{iQ,{\bar{c}}^{\ a}\},
\end{eqnarray}
where $Q$ is the BRS-charge $Q=\int d^3x\{B^a(D_{\mu}^{ab}c^b)-{\partial}_0B^ac^a+i\frac{g}{2}
f^{abc}{\partial}_0{\bar{c}}^{\ a}\ c^bc^c\}$. (see [7] for details) From (31), it is seen that 
$\delta F_{\mu\nu}^a=gf^{acb}F_{\mu\nu}^c c^b$ and the invariance of (28) under (31) can be verified.

\vspace{0.5cm}

Though the local gauge invariance is explicitly broken by the gauge fixing, (28) has global gauge symmetry.
This global gauge transformation (Second Global Transformation) 
\begin{eqnarray}
\Delta A_{\mu}^a&=&f^{abc}{\theta}^bA_{\mu}^c, \nonumber \\
\Delta {\psi}_i&=&-i(t^a)_{ij}{\theta}^a\ {\psi}_j, \nonumber \\
\Delta {\bar{\psi}}_i&=&i{\bar{\psi}}_j(t^a)_{ji}{\theta}^a, \nonumber \\
\Delta B^a&=& f^{abc}{\theta}^bB^c, \nonumber \\
\Delta c^a&=&f^{abc}{\theta}^bc^c, \nonumber \\
\Delta {\bar{c}}^{\ a}&=&f^{abc}{\theta}^b{\bar{c}}^{\ c}, 
\end{eqnarray}
where ${\theta}^a$ is the global gauge parameter, generates the conserved Noether current 
\begin{eqnarray}
{\cal{J}}_{\mu}^a&=&f^{abc}A^{\nu b}F_{\nu\mu}^c+j_{\mu}^a+f^{abc}A_{\mu}^bB^c-if^{abc}{\bar{c}}^{\ b}
(D_{\mu}^{cd}c^d)+if^{abc}{\partial}_{\mu}{\bar{c}}^{\ b}\ c^c, \nonumber \\
&=&J_{\mu}^a+f^{abc}A_{\mu}^bB^c-if^{abc}{\bar{c}}^{\ b}(D_{\mu}^{cd}c^d)+if^{abc}({\partial}_{\mu} 
{\bar{c}}^{\ b})c^c,
\end{eqnarray} 
where in the last step we used the third relation in (27). 

\vspace{0.5cm}

We now  consider the first equation in (30) and rewrite that as 
\begin{eqnarray}
{\partial}_{\mu}F^{\mu\nu a}+gf^{acb}A_{\mu}^c F^{\mu\nu b}&=&{\partial}^{\nu}B^a-gj^{\nu a}
-igf^{abc}({\partial}^{\nu}{\bar{c}}^{\ b})c^c.
\end{eqnarray}
This, in view of (33) can be written as 
\begin{eqnarray}
{\partial}_{\mu}F^{\mu\nu a}+g{\cal{J}}^{\nu a}&=&(D^{\nu ac}B^c)-igf^{abc}{\bar{c}}^{\ b}(D^{\nu cd}
c^d).
\end{eqnarray}
The right side of (35) can be expressed, using the BRS transformations (31), as 
$-i{\delta}(D^{\nu ab}{\bar{c}}^{\ b})$ and so (35) becomes 
\begin{eqnarray}
{\partial}_{\mu}F^{\mu\nu a}+g{\cal{J}}^{\nu a}&=&\{Q,D^{\nu ab}{\bar{c}}^{\ b}\}.
\end{eqnarray}
This quantum equation of motion is to be compared with the classical equation of motion (27). We note that 
$J^{\nu a}$ in (27) is replaced by ${\cal{J}}^{\nu a}$ in (36) and the right side is expressed as a 
BRS-variation. Both $J^{\nu a}$ and ${\cal{J}}^{\nu a}$ are ordinarily conserved. That the quantum 
equation (34) can be written in the form (36) was first shown by Ojima [9]. 

\vspace{0.5cm}

The vector space for the non-Abelian gauge fields, on which the quantum equations act is an indefinite 
metric space. A physical subspace of this is 
to be defined. It was shown by Kugo and Ojima [10] that the physical space is defined by the condition 
\begin{eqnarray}
Q|\phi\rangle &=&0.
\end{eqnarray}
Taking the expectation value of (36) between the physical states and using (37) it follows 
\begin{eqnarray}
\langle \phi|{\partial}_{\mu}F^{\mu\nu a}+g{\cal{J}}^{\nu a}|\phi \rangle &=&0.
\end{eqnarray}
This expression when compared with the classical equation of motion (27) shows that the Ehrenfest theorem
is not fully satisfied. The global conserved current ${\cal{J}}^{\nu a}$ differs from the conserved 
current $J^{\nu a}$, as seen from (33). Now we consider (33) and note that this difference is given by 
$f^{abc}A_{\mu}^bB^c-if^{abc}{\bar{c}}^{\ b}(D_{\mu}^{cd}c^d)+if^{abc}({\partial}_{\mu}{\bar{c}}^{\ b})c^c$.
The first two terms can be expressed using (31) as $\delta (if^{abc}{\bar{c}}^bA_{\mu}^c)$ noting that 
when the BRS-variation crosses the ghost field it picks up a sign. So the first two terms can be rewritten 
as 
$\{-Q,f^{abc}{\bar{c}}^{\ b}A_{\mu}^c\}$ and this when taken between the physical states vanishes. Then, 
(38) becomes 
\begin{eqnarray}
\langle \phi|{\partial}_{\mu}F^{\mu\nu a}+gJ^{\nu a}+if^{abc}({\partial}^{\nu}{\bar{c}}^{\ b})c^c|\phi 
\rangle &=&0.
\end{eqnarray}
This still differs from the classical equation of motion by a term involving ghosts only. 

\vspace{0.5cm}

We now take up the quantum lagrangian (28) and note it is invariant under the scale 
transformation (Third Global Transformation) 
\begin{eqnarray}
c^a\rightarrow e^{\alpha}c^a&;&{\bar{c}}^{\ a}\rightarrow e^{-\alpha}{\bar{c}}^{\ a},
\end{eqnarray}
with $\alpha$ a constant. This global transformation affects only the FP-ghost fields in (28). The Noether 
current corresponding to this transformation is given by 
\begin{eqnarray}
J_{gh}^{\lambda}&=&{\delta}_{\alpha}c^a\ \frac{\partial {\cal{L}}}{\partial({\partial}_{\lambda}c^a)}+
{\delta}_{\alpha}{\bar{c}}^{\ a}\ \frac{\partial {\cal{L}}}{\partial({\partial}_{\lambda}{\bar{c}}^{\ a})}, 
\nonumber \\
&=&i{\bar{c}}^{\ a}(D^{\lambda ab}c^b)-i({\partial}^{\lambda}{\bar{c}}^{\ a})c^a,
\end{eqnarray}
as $\alpha$ is arbitrary.  The corresponding conserved 
charge $Q_{gh}\ =\ (Q_{gh})^{\dagger}$ is called the FP-ghost charge generating the above scale 
transformation on the ghost fields, leaving other fields invariant [7]. This is given by 
\begin{eqnarray}
{\delta}_{gh}c^a=[iQ_{gh},c^a]\ =\ c^a&;&{\delta}_{gh}{\bar{c}}^{\ a}=[iQ_{gh},{\bar{c}}^{\ a}]\ =\ 
-{\bar{c}}^{\ a}.
\end{eqnarray}
Using the above, the third term in (39) can be written as 
\begin{eqnarray}
if^{abc}({\partial}_{\mu}{\bar{c}}^{\ b})c^c&=&-\frac{1}{2}{\delta}_{gh}(if^{abc}({\partial}_{\mu}
{\bar{c}}^{\ b})c^c), \nonumber \\
&=&\frac{1}{2}[Q_{gh},f^{abc}({\partial}_{\mu}{\bar{c}}^{\ b})c^c],
\end{eqnarray}
as ${\delta}_{gh}$ when crosses a FP-ghost field picks up a sign. 

\vspace{0.5cm}

We defined the physical subspace in (37) as the assembly of states in the indefinite metric Hilbert space 
annihilated by the BRS-Charge. We now restrict the physical subspace further by another subsidiary 
condition 
\begin{eqnarray}
Q_{gh}|\phi \rangle &=&0.
\end{eqnarray}
Then, using (43) in the last term in (39) and in view of the further restriction (44) on the physical 
states, (39) becomes 
\begin{eqnarray}
\langle \phi |{\partial}_{\mu}F^{\mu\nu a}+gJ^{\nu a}|\phi \rangle &=&0,
\end{eqnarray}
showing that the expectation value of the quantum equation of motion for the non-Abelian gauge fields 
agrees with the classical equation of motion (27). 

\vspace{0.5cm}

Now we examine the other quantum equations of motion 
in (30). The second equation in (30), in view of the BRS-transformation (31) can be written as 
${\partial}_{\mu}A^{\mu a}+\alpha \{Q,{\bar{c}}^{\ a}\}=0$ which when its expectation value between the 
physical states defined in (37) are taken gives $\langle \phi|{\partial}_{\mu}A^{\mu a}|\phi \rangle =0$,
giving the gauge fixing condition. The third equation in (30), in view of the third global transformation 
(42), is written as $[iQ_{gh},(D_{\mu}^{ab}({\partial}^{\mu}{\bar{c}}^{\ b}))]=0$ and its expectation value 
taken between the physical states vanishes on account of (44). The fourth equation in (30), using the 
BRS-transformation (first global transformation) becomes $[iQ,{\partial}_{\mu}A^{\mu a}]$ whose 
expectation value between the physical states vanishes on account of (37). This shows the validity of 
Ehrenfest's theorem for the quantum non-Abelian theory. We have made use of three global transformations to 
arrive at this conclusion. 

\vspace{1.0cm}

{\noindent{\bf{4. Summary}}}

\vspace{0.5cm}

The Ehrenfest theorem in quantum mechanics is shown to be satisfied in the quantum field theory by 
suitably taking the physical subspace for the gauge fields. In the Abelian quantum field theory, 
the one subsidiary condition on the physical states of the photon is enough to show this.
In the case of non-Abelian field theory, the subsidiary condition (37) is not enough and one has to 
further restrict the physical space by (44). Then the expectation value of the quantum equations of 
motion between the physical states satisfying (37) and (44) agree with the classical equations of 
motion, including the gauge fixing condition. 

\vspace{1.0cm}

{\noindent{\bf{References}}}

\vspace{1.0cm}

\begin{enumerate}
\item P.Ehrenfest, Zeits. f. Physik, {\bf{45}} (1927) 455.
\item L.I.Schiff, {\it{Quantum Mechanics}}, McGraw-Hill. Second Edition, 1955. 
\item J.D.Bjorken and S.D.Drell, {\it{Relativistic Quantum Fields}}, McGraw-Hill, 1965.
\item S.N.Gupta, Proc. Phys. Soc. {\bf{A63}} (1950) 681; \\
      K.Bleuler, Helv. Phys. Acta. {\bf{23}} (1950) 567. 
\item P.M.Mathews, M.Seetharaman and M.T.Simon, Phys.Rev. {\bf{D9}} (1974) 1700.
\item N.Nakanishi, Prog. Theor. Phys. {\bf{35}} (1966) 1111. 
\item N.Nakanishi and I.Ojima, {\it{Covariant Operator Formalism of Gauge Theories and Quantum Gravity}}, 
      World Scientific, 1990; L.H.Ryder, {\it{Quantum Field Theory}}, Cambridge University Press, 1996.
\item C.Becchi, A.Rouet and R.Stora, Ann. Phys. {\bf{98}} (1976) 287.
\item I.Ojima, Nucl. Phys. {\bf{B143}} (1978) 340. 
\item T.Kugo and I.Ojima, Phys. Lett. {\bf{73B}} (1978) 459. \\
      T.Kugo and I.Ojima, Prog. Theor. Phys. {\bf{60}} (1978) 1869. 
\end{enumerate}       

\end{document}